\documentclass[lettersize,journal]{IEEEtran}
\usepackage{amsmath,amsfonts}
\usepackage{algorithmic}
\usepackage{algorithm}
\usepackage{array}
\usepackage[caption=false,font=normalsize,labelfont=sf,textfont=sf]{subfig}
\usepackage{textcomp}
\usepackage{stfloats}
\usepackage{url}
\usepackage{verbatim}
\usepackage{graphicx}
\usepackage{cite}
\usepackage{booktabs}
\usepackage{amssymb}
\usepackage{utfsym}
\usepackage{fontawesome}
\usepackage[utf8]{inputenc}
\usepackage{url}
\usepackage{bbding}
\usepackage{pifont}
\usepackage{wasysym}
\usepackage{color}
\usepackage{bm}
\definecolor{blue}{RGB}{0,112,192}
\definecolor{green}{RGB}{146,208,80}
\definecolor{gray}{RGB}{127,127,127}
\definecolor{vis-green}{RGB}{61,145,64}
\definecolor{red}{RGB}{202,0,22}
\definecolor{yellow}{RGB}{255,192,0}

\usepackage{newfloat}
\usepackage{listings}

\begin{document}
\title{EnchantDance: Unveiling the Potential of \\ Music-Driven Dance Generation}
\author{Bo Han\textsuperscript{1}, Teng Zhang\textsuperscript{1}, Zeyu
  Ling\textsuperscript{1}, Feilin Han\textsuperscript{2,\dag}
    \thanks{\dag Corresponding author.}
    \thanks{\textsuperscript{1} Zhejiang University \textsuperscript{2} Tongji University}
    }

\maketitle
\begin{abstract}
The task of music-driven dance generation involves creating coherent dance movements that correspond to the given music. While existing methods can produce physically plausible dances, they often struggle to generalize to out-of-set data. The challenge arises from three aspects: 1) the high diversity of dance movements and significant differences in the distribution of music modalities, which make it difficult to generate music-aligned dance movements. 2) the lack of a large-scale music-dance dataset, which hinders the generation of generalized dance movements from music. 3) The protracted nature of dance movements poses a challenge to the maintenance of a consistent dance style. In this work, we introduce the EnchantDance framework, a state-of-the-art method for dance generation. Due to the redundancy of the original dance sequence along the time axis, EnchantDance first constructs a strong dance latent space and then trains a dance diffusion model on the dance latent space. To address the data gap, we construct a large-scale music-dance dataset, ChoreoSpectrum3D, which includes four dance genres and has a total duration of 70.32 hours, making it the largest reported music-dance dataset to date. To enhance consistency between music genre and dance style, we pre-train a music genre prediction network using transfer learning and incorporate music genre as extra conditional information in the training of the dance diffusion model. Extensive experiments demonstrate that our proposed framework achieves state-of-the-art performance on dance quality, diversity, and consistency.
\end{abstract}


\section{Introduction}
Music and dance, as expressive art forms, have captivated human emotions throughout history and find wide application in movies, games, and various industries in modern society~\cite{hanna1987dance}. The creation of high-quality 3D dance animation, however, remains a costly endeavor, demanding the involvement of experienced dancers and expensive motion-capture equipment. The utilization of computational methods for automatic dance generation can alleviate the burdensome process of manual creation. Such methods~\cite{chen2021choreomaster,siyao2022bailando,tseng2023edge} have the potential to aid animators in creating novel dance sequences from music. 

Despite these advancements in automatic choreography, achieving high-fidelity music-driven dance motion generation remains challenging due to the intricate interplay between motion diversity and acoustic complexity. 
\textbf{Weak correlation between music and dance.} The diversity of dance movements, which shapes the high-dimensional distribution of body movements, and the rich genre and beat information in music lead to a one-to-many mapping problem. To ensure the generalization of the model, a large-scale training dataset is often required. \textbf{Scarcity of large-scale data.} Compared to the text-to-motion dataset~\cite{guo2022generating}, which can span dozens of hours, the music-to-dance field lacks sufficient data volume to generate high-fidelity and well-generalized dance movements. \textbf{Fragile consistency of dance genres.} Dance styles are characterized by distinct movement patterns, with specific constraints on the suitability and relevance of dance poses to a given genre and musical accompaniment. The absence of genre-specific information may result in the incorporation of multiple dance styles within a single piece, thereby diminishing the expressiveness and coherence of the performance (e.g., Latin dance motions in ballet music). However, dance sequences are often long-term, and it will become difficult to maintain the consistency of dance genres.

\begin{figure}[t]
    \centering
    \includegraphics[width=\columnwidth]{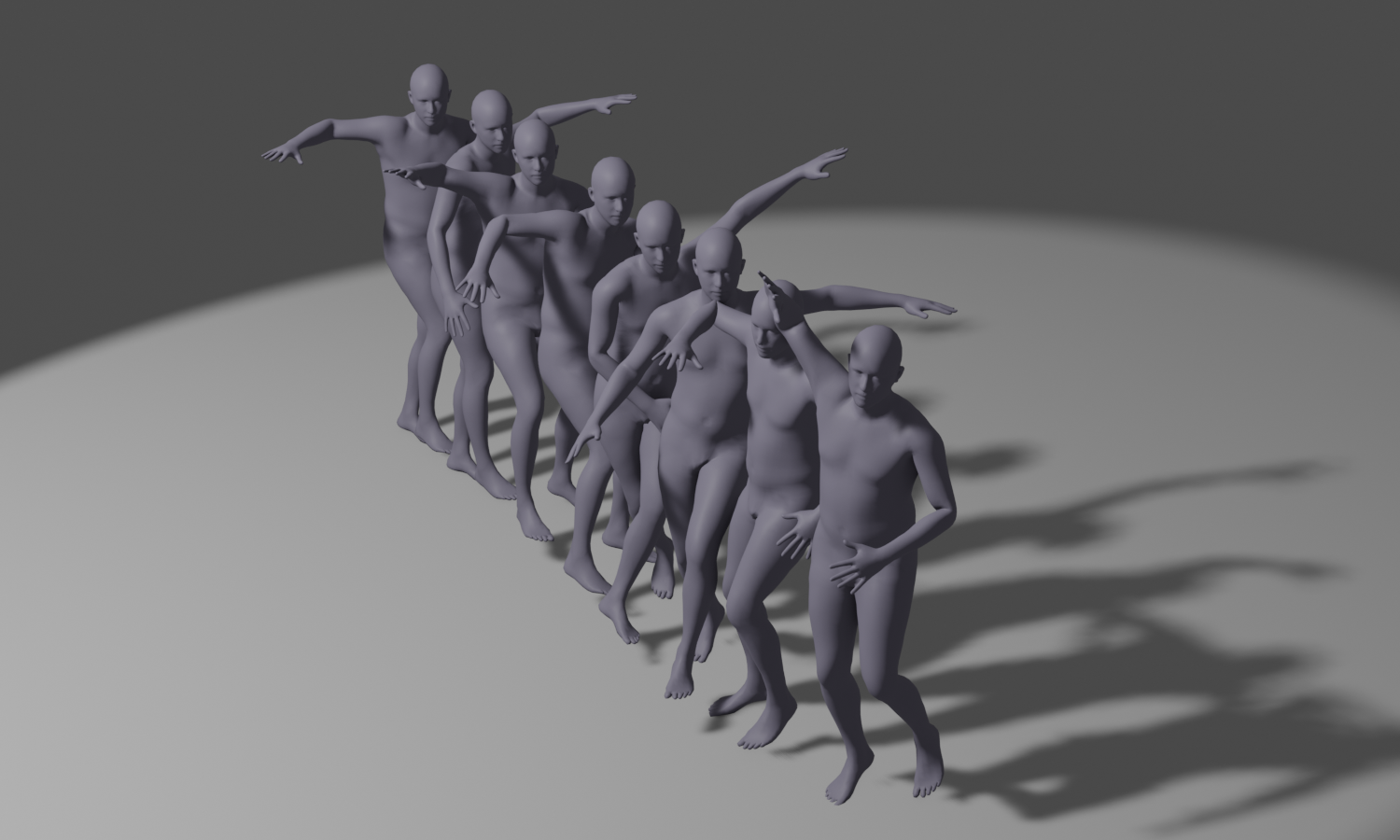}
    \caption{Visualization of a generated Latin dance sequence. EnchantDance achieves high-fidelity motion synthesis with natural body coordination, accurately reflecting the passionate and energetic style of the input music while maintaining long-term stability.}
    \label{fig:teaser}
\end{figure}

Current genre-oriented dance generation frameworks frequently employ manually appending genre tags during the inference stage. Given the strong correlation between dance styles and music genres, GTN-Bailando~\cite{zhuang2023gtn} introduces a Genre Token Network that infers genres from music to enhance the consistency of dance generation. However, constrained by the scarcity of large-scale music-dance data, the model continues to exhibit poor generalization capability. We design a simple yet effective music genre prediction network leveraging transfer learning. Specifically, we have incorporated a simple CNN block into the Audio Spectrogram Transformer (AST) pre-trained on the ImageNet and AudioSet~\cite{hershey2017cnn} to facilitate music genre prediction. This approach enables music genre prediction with a minimal number of training parameters, obviating modifications to the entire model and enhancing generalization. 

Existing 3D music and dance datasets are generally limited in duration, which adversely impacts the performance of models when applied to out-of-set data. The widely used AIST++ dataset~\cite{li2021ai} has a duration of only 5.19 hours. In comparison, the HumanML3D dataset~\cite{guo2022generating} for the text-to-motion field has a duration of 28.59 hours, highlighting the need for a large-scale, high-quality dance dataset. To address this need, we present the ChoreoSpectrum3D Dataset, which covers four coarse-grained genres of dance types, with a total duration of 70.32 hours.

In this paper, we initially devise a variational autoencoder founded on the Transformer architecture to compress the dance motion sequence into the latent space, effectively eliminating redundancy in the original dance sequence. Subsequently, we propose a diffusion model predicated on dance latent space to learn a robust mapping from the music conditional distribution to the dance latent vector. This approach not only facilitates the generation of reasonable and diverse dance motion sequences but also eliminates redundant motion information and noise while mitigating the issue of slow sampling and poor generalization. Extensive experiments on the AIST++ and our ChoreoSpectrum3D datasets demonstrate the superiority of our method with respect to dance generation quality, diversity, and music-dance consistency. Concurrently, comprehensive out-of-set data evaluation has corroborated the generalization capability of our method and the efficacy of the dataset. The code, ChoreoSpectrum3D dataset, and demos can be found in the Supplementary Materials.

\begin{table*}[htp]
\small
\centering
\caption{3D Dance Datasets Comparisons} 
\label{tab:dataset}
\begin{tabular}{c|c|c|c|c|c|c|c}
\toprule
Dataset  & $3\text{D Jointpos}_{\text{pos}}$ & $3\text{D Jointrot}_{\text{rot}}$ & $2\text{D Keypoints}$ & Genres & Fbx & Hours & Available \\ \midrule
GrooveNet~\cite{alemi2017groovenet}  & $\usym{2713}$ & $\usym{2717}$ & $\usym{2717}$ & 1 & $\usym{2717}$ & 0.38 & $\usym{2717}$ \\
Dance w/Melody~\cite{tang2018dance}  & $\usym{2713}$ & $\usym{2717}$ & $\usym{2717}$ & 4 & $\usym{2717}$ & 1.57 & $\usym{2713}$ \\
DanceNet~\cite{zhuang2022music2dance}  & $\usym{2713}$ & $\usym{2717}$ & $\usym{2717}$ & 2 & $\usym{2717}$ & 0.96 & $\usym{2717}$ \\
DeepDance~\cite{sun2020deepdance}  & $\usym{2713}$ & $\usym{2713}$ & $\usym{2717}$ & 4 & $\usym{2717}$ & 0.35 & $\usym{2713}$ \\
PhantomDance~\cite{li2022danceformer}  & $\usym{2713}$ & $\usym{2717}$ & $\usym{2717}$ & 13 & $\usym{2717}$ & 9.63 & $\usym{2717}$ \\ 
AIST++~\cite{li2021ai}  & $\usym{2713}$ & $\usym{2713}$ & $\usym{2713}$ & 10 & $\usym{2717}$ & 5.19 & $\usym{2713}$ \\ 
MMD~\cite{chen2021choreomaster}  & $\usym{2713}$ & $\usym{2713}$ & $\usym{2717}$ & 4 & $\usym{2717}$ & 9.91 & \includegraphics[width=0.7em]{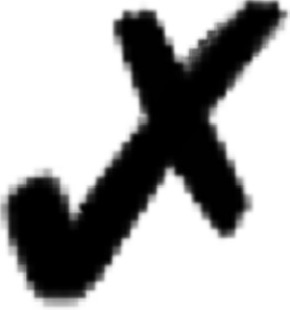}\\ 
FineDance~\cite{Li_2023_ICCV}  & $\usym{2713}$ & $\usym{2713}$ & $\usym{2717}$ & 22 & $\usym{2713}$ & 14.6 & \includegraphics[width=0.7em]{PIC/semicheck.jpg} \\ \midrule
ChoreoSpectrum3D  & $\usym{2713}$ & $\usym{2713}$ & $\usym{2713}$ & 4 & $\usym{2713}$ & $\textbf{70.32}$ & $\usym{2713}$ \\
\bottomrule 
\multicolumn{8}{r}{
  \small \textit{Note:} $\usym{2713}$, \includegraphics[width=0.7em]{PIC/semicheck.jpg}, and $\usym{2717}$ denote full, partial, and not available, respectively.
} 
\end{tabular}
\end{table*}

\section{Related Work}

\subsection{Learning-based Dance Generation}
Early methods~\cite{kim2003rhythmic,alemi2017groovenet,tang2018dance} for music-driven dance generation typically employ motion graphs to splice dance segments and form complete dance movements. However, these methods are limited to creative dances. Autoregressive models~\cite{li2021ai,siyao2022bailando} have also been used for music-driven generative dance, but often suffer from limb drift and poor generalization. 

The diffusion model exhibits robust vitality and is well-suited for the generation of data characterized by high diversity. Typically, EDGE~\cite{tseng2023edge} employs a Transformer-based diffusion model that adheres to the DDPM~\cite{ho2020denoising} in defining the diffusion process, thereby enabling direct prediction of the original dance sequence. The model exhibits the capability to generate physically plausible dances that are consistent with the input music. 
However, raw motion sequences exhibit redundancy along the time axis, and diffusion models operating on raw sequence data typically incur significant computational overhead during both training and inference stages~\cite{rasul2021autoregressive,ho2022imagen}, rendering them inefficient. Moreover, raw motion data is susceptible to noise contamination, which may prompt strong diffusion models to learn cues for probability mapping from conditional inputs to noisy motion sequences, resulting in the production of artifacts. 

\subsection{Generative Models for Dance Generation}
Numerous early studies~\cite{ofli2011learn2dance,lee2013music,fan2011example} have employed motion retrieval paradigms to synthesize dance movements. This approach involves segmenting a given music piece and retrieving corresponding dance clips, which are then integrated to generate a complete dance sequence but tend to create unrealistic dances that lack creativity. Recently, generative models such as Generative Adversarial Networks (GANs)~\cite{goodfellow2020generative}, Variational Autoencoders (VAEs)~\cite{kingma2013auto}, and Diffusion models~\cite{ho2020denoising} have been successfully applied to various data modalities, including images~\cite{pan2023drag}, text~\cite{touvron2023llama}, audio~\cite{huang2023make}, and beat conditions~\cite{huang2024beat}. 
Consequently, numerous studies have utilized these generative models to tackle the challenge of music-driven dance motion generation~\cite{zhuang2022music2dance,huang2020dance}. For instance, the Full-Attention Cross-Model Transformer (FACT) framework~\cite{li2021ai} has been adopted to generate dance movements from music. Similarly, Bailando~\cite{siyao2022bailando} utilized a choreographic codebook and the Transformer framework to match these dance units with music. 

Despite the ability of autoregressive models to generate music-consistent motions, these methods are limited in limb drift and poor generalization to out-of-set data. Recent studies have explored more flexible generative paradigms. DanceEditor~\cite{zhang2025danceeditor} introduces an iterative editing framework with open-vocabulary descriptions, and Lodge++ \cite{li2024lodge++} proposes a two-stage coarse-to-fine strategy to achieve high-quality music-to-dance mapping. Among these advancements, the diffusion-based approach, EDGE \cite{tseng2023edge}, incorporates Jukebox music features, outperforming in the music and dance consistency. Building upon this, more genre-aware approaches have emerged to enhance stylistic diversity. MEGADance \cite{yang2025megadance} further enhances stylistic control by utilizing a Mixture-of-Experts (MoE) architecture, and GEN~\cite{li2024exploring} proposed a multi-modal control framework with genre control, semantic control, and spatial control. However, the implementation of these methods on raw dance sequences impedes generation quality and generalization capability due to the presence of redundant information and noise in the original dance sequences.

We initially compress the raw action sequence using a variational autoencoder to eliminate redundant information, followed by the implementation of a diffusion model on the dance latent space. This approach significantly enhances the generalization capability of the model when applied to out-of-set data. Additionally, we have incorporated supplementary music genre information to ensure consistency between dance move styles and music genres.

\subsection{3D Music-Dance Dataset}
AIST++ dataset~\cite{li2021ai} provides 3D music and dance data for 10 genres. However, there is a high degree of overlap between music and dance movements, with one piece of music corresponding to multiple dances. Additionally, the dataset only contains 5.19 hours of motion data. The GrooveNet dataset~\cite{alemi2017groovenet} utilizes motion capture equipment to obtain dance motion but is limited to electronic dance music and only lasts 23 minutes. The Dance with Melody dataset~\cite{tang2018dance} contains four dance genres, but the motion duration is only 1.57 hours and the dance movements lack variety and have low matching with the music. The Music2Dance dataset~\cite{zhuang2022music2dance} claims to contain two dance genres, modern dance and curtilage dance, with precise alignment with the music, but it only lasts for 1 hour and is not open source. The MMD dataset~\cite{chen2021choreomaster} and FineDance dataset~\cite{Li_2023_ICCV} contain 9.91 and 14.6 hours of dance sequences, respectively; however, they are not fully accessible, with only 1.48 and 7.7 hours being currently available for use. Compared to the tens of hours of data in the public datasets HumanML3D~\cite{guo2022generating} and HumanLong3D~\cite{han2023amd} in the text-to-motion field, it still lacks scale and is insufficient for fully developing this task. 
As a result, we construct ChoreoSpectrum3D Dataset with a duration of approximately 70.32 hours. Additionally, it includes precise SMPL pose and joint parameters for dance movements, as well as aligned music data. We analyze datasets from multiple perspectives, including genres, joints, duration, and availability. A comparison between ChoreoSpectrum3D and existing 3D dance datasets is presented in Table~\ref{tab:dataset}. 

\begin{figure*}[t] 
    \centering
    \setlength{\tabcolsep}{1pt}
    
    \begin{tabular}{cccc}
        \includegraphics[width=0.245\textwidth]{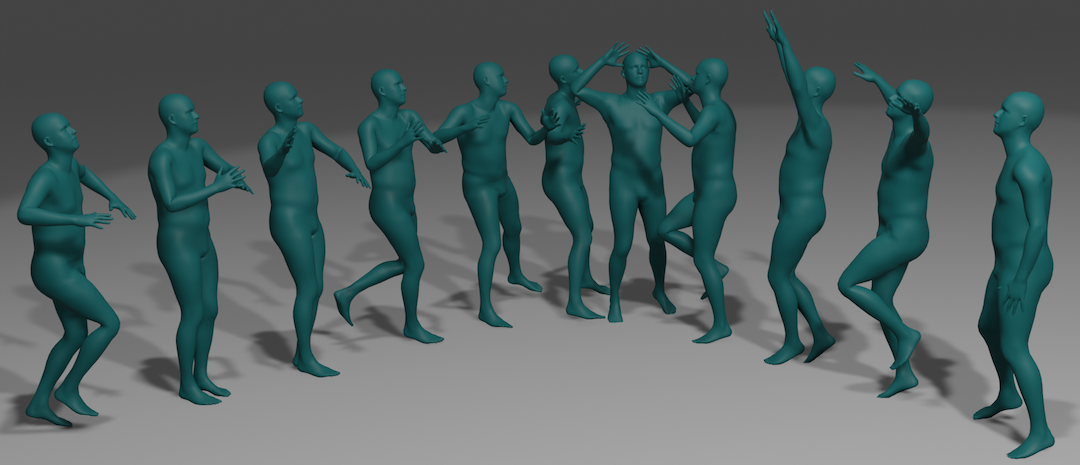} &
        \includegraphics[width=0.245\textwidth]{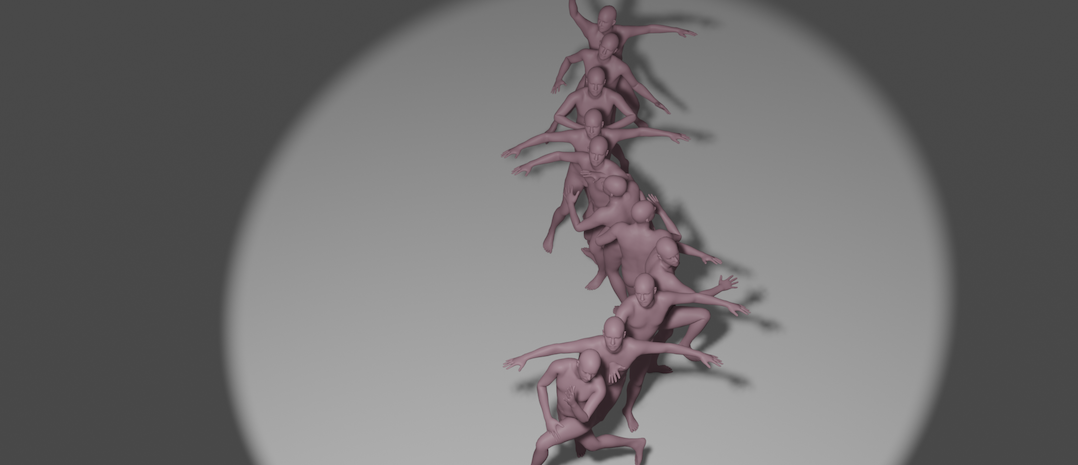} &
        \includegraphics[width=0.245\textwidth]{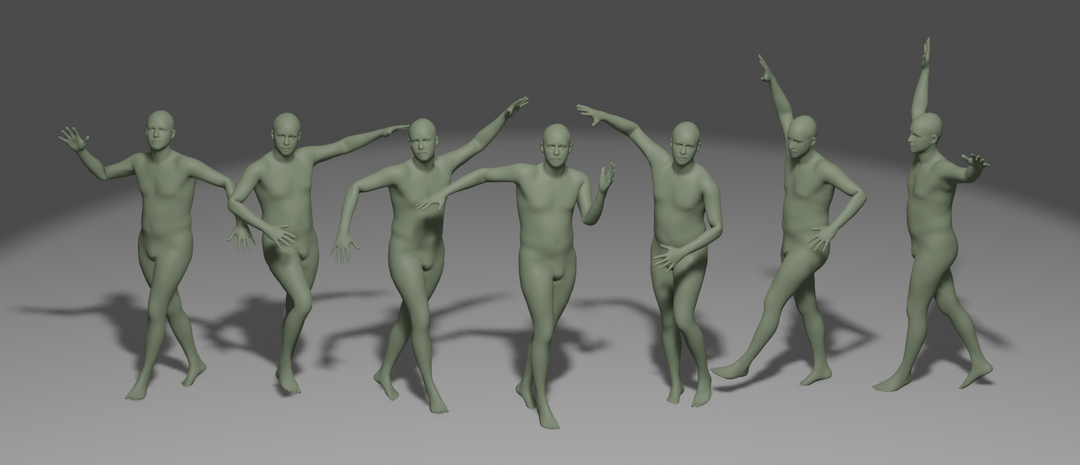} &
        \includegraphics[width=0.245\textwidth]{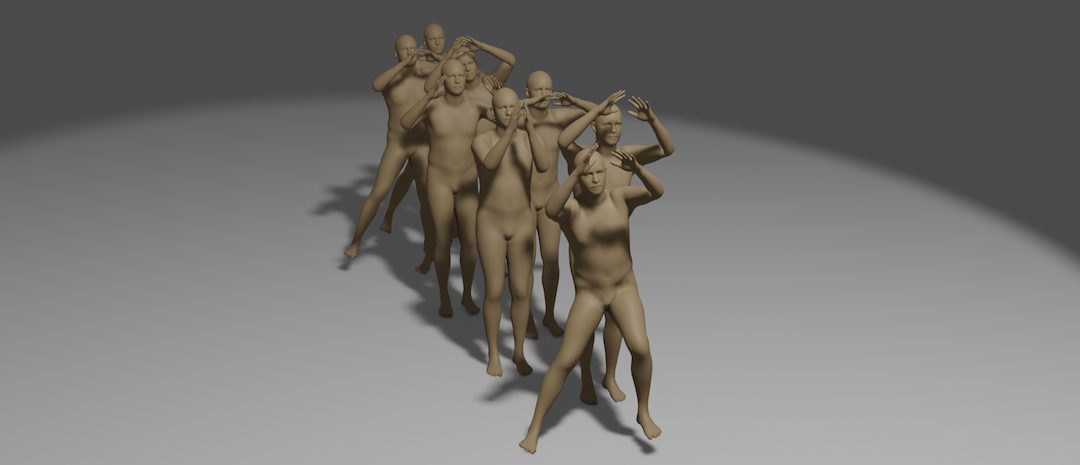} \\
        
        \makebox[0.245\textwidth][c]{\small (a) Pop} & 
        \makebox[0.245\textwidth][c]{\small (b) Ballet} & 
        \makebox[0.245\textwidth][c]{\small (c) Latin} & 
        \makebox[0.245\textwidth][c]{\small (d) House} \\
    \end{tabular}
    \caption{Representative samples from the ChoreoSpectrum3D dataset, including four major dance genres.}
    \label{fig:dataset_samples}
\end{figure*}

\section{ChoreoSpectrum3D Dataset}
While existing works~\cite{tang2018dance,li2022danceformer,sun2020deepdance,chen2021choreomaster, Li_2023_ICCV} have reported music-dance datasets, the availability of large-scale choreography datasets remains limited. Currently, the widely utilized dataset for music-driven dance generation tasks is the AIST+++ dataset~\cite{li2021ai}. This dataset adequately caters to the requirements of dance generation tasks. Nevertheless, the limited duration of the dataset poses a challenge as the distribution space of dance moves remains considerably large. This limitation hinders the ability of existing models to generalize effectively and generate high-quality dance moves. It emphasizes the need for additional data or augmentation.
Hence, we construct a choreography-oriented large-scale dataset spanning a duration of 70.32 hours.

\subsection{Dance Genre Categories}

The dance genres in the AIST++ dataset include Break, Pop, Lock, Middle Hip-hop, LA style Hip-hop, Waack, Krump, Street Jazz, Ballet Jazz, and House. Based on the suggestions of professional musicians and dancers, we re-classified them according to their music styles. To align our dance data more closely with the standards of dance artists, under the guidance of professional dancers, we classify dances into four genre-based categories: pop (Breaking, Locking, Hip-hop, Popping, Urban, and Jazz), ballet, Latin, and House dance. Their corresponding music styles are also distinct from one another and suitable for music genre prediction. The examples are shown in the Figure~\ref{fig:dataset_samples}.

\subsection{Motion Annotation}
Motion capture equipment can provide clear and consistent dance movements, but it is often prohibitively expensive. As an alternative, we source dance videos of professional performers from the internet and estimate human body parameters from these videos. We employ the 3D human motion estimation model~\cite{li2022cliff,goel2023humans} to obtain the SMPL motion parameters of characters in videos. However, we observe that dance videos often contain invalid motion frames. To address this issue, we design an automatic filtering method to eliminate implausible motion frames. In an effort to preserve the fluidity of the dance sequence, we retain only those continuous dance clips comprising 200 or more valid frames. 
Concurrently, to enhance the versatility of our dataset, we have also generated detailed annotations with reference to the HumanML3D format, widely employed in the text-to-motion domain. This primarily encompasses the acquisition of forward kinematic joint coordinates, extraction of kinematic vector representation, and computation of sample mean-variance. Further details about the ChoreoSpectrum3D dataset can be found in the Supplementary Material.

\subsection{Music features}
We not only address the Jukebox~\cite{dhariwal2020jukebox} music feature, which has previously exhibited robust performance on music-specific tasks~\cite{tseng2023edge}, but also incorporate audio spectrum features extracted by the public audio processing toolbox Librosa~\cite{jin2017towards}, including mel frequency cepstral coefficients (MFCC), MFCC delta, constant-Q chromagram, tempogram, and onset strength.

\section{EnchantDance}

\begin{figure*}[htb]
    \centering
    \includegraphics[width=0.95\linewidth]{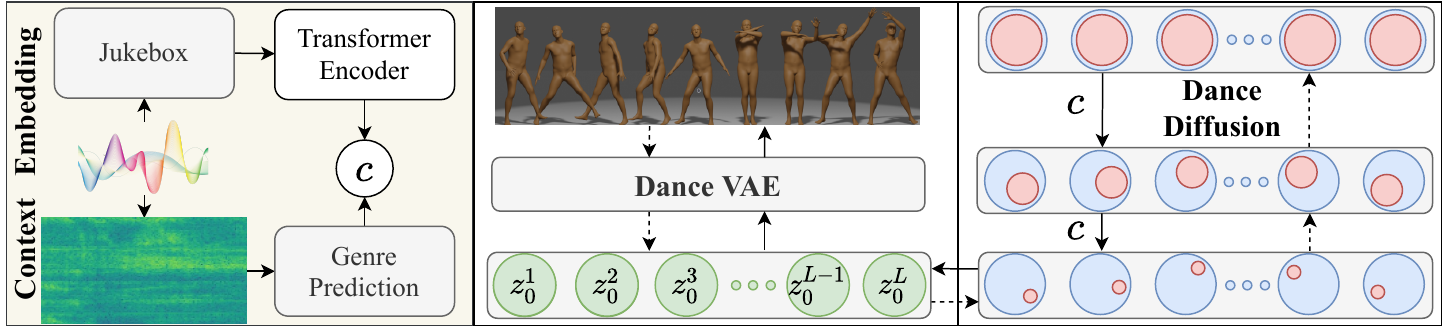}
    \caption{Overview of the EnchantDance Framework. Upon completion of pre-training for the Genre Prediction Network and Dance VAE, their weights are frozen (gray). The Dance Diffusion comprises a forward process (dotted line) and a denoising process (solid line). The forward process incrementally introduces noise to the isotropic Gaussian distribution, while the denoising process progressively removes noise from the distribution of dance movements. $c$ represents context embedding.}
    \label{fig:overall}
\end{figure*}

\subsection{Overview}
Unlike semantic-heavy text-to-motion tasks, music-to-dance synthesis necessitates sophisticated parsing of physical signals. EnchantDance addresses this challenge by synergizing low-level Jukebox features for melodic guidance with high-level genre-aware features for contextual consistency. Our genre prediction network, inspired by LoRA-style fine-tuning, adapts pre-trained AST representations through CNN-based skip connections to extract a high-dimensional stylistic vector. This latent-based approach allows us to push the performance boundaries established by existing diffusion models while maintaining stylistic integrity.

Formally, given a piece of music $M$. To ensure the alignment between the generated dance motions and the emotions conveyed in the music, we first pre-train the Genre Prediction Network to obtain genre categories $\hat{g} \in G$, where $G$ is the set of pre-defined genre categories. Subsequently, we concatenate the genre category $\hat{g}$ with the music features $\overline{M} \in \mathbb{R}^{T \times C}$,  where $T$ denotes that the frame number of the action is aligned with the music, while $C$ represents the dimension of the feature. These concatenated features are then utilized as conditional information for the generation of dance movements $X \in \mathbb{R}^{T \times S}$, where $S$ represents the dimension of the SMPL parameter.

The architecture of the proposed network is depicted in Figure~\ref{fig:overall} and comprises three primary modules: Context Embedding module for multi-level music parsing, DanceVAE for latent representation, and Dance Diffusion for motion synthesis. The Context Embedding module employs Jukebox~\cite{dhariwal2020jukebox} to extract music features. The DanceVAE extracts latent features of dance movements, followed by training the Dance Diffusion model in the latent space.

\subsection{Dance VAE}
The ChoreoSpectrum3D dataset, much larger than AIST++, presents a vast and diverse motion space. Unlike methods such as EDGE~\cite{tseng2023edge} that perform diffusion directly on raw motion data—which often involves significant redundancy—our approach emphasizes the necessity of motion compression. As observed in Bailando~\cite{siyao2022bailando}, compressing dance motions into dense latent features not only reduces redundancy but also significantly enhances generative quality. Therefore, we adopt a VAE-based architecture to project motions into a compact latent space before the diffusion process.

To eliminate temporal redundant information in the original dance sequence, we develop the dance variational autoencoder (VAE) based on the Transformer architecture~\cite{vaswani2017attention}, which comprises a Transformer encoder $\mathcal{E}$ and a Transformer decoder $\mathcal{D}$. The encoder constructs a low-dimensional, high-density latent space, while the decoder reconstructs the original dance sequence by upsampling the latent space. 
The input to the dance VAE is the SMPL parameter, and the output is the reconstructed SMPL parameter. 
Additionally, inspired by UNet networks, we establish a long skip connection between the two Transformer architectures to ensure temporal correlation of the motions. The training objective of the model is to minimize the reconstruction loss and Kullback-Leibler (KL) loss:

\begin{equation}
    \label{equ:VAE-loss}
    L=\min_{\mathcal{E},\mathcal{D}}\max_{\mathcal{N}}L_{rec}(x,\mathcal{D}(\mathcal{E}(x)))+L_{KL}
\end{equation}

\begin{equation}
    \label{equ:rec-loss}
    L_{rec}=\sum_{n=1}^{N}\Vert{x-\mathcal{D}(\mathcal{E}(x))}\Vert^2
\end{equation}

\begin{equation}
    \label{equ:KL-loss}
    L_{KL}=\log(\mathcal{N}(x))+\log(1-\mathcal{N}(\tilde{x}))
\end{equation}

The dance encoder embeds the frame-wise SMPL parameters $x \in X$ as input into the latent space. The dance latent space is characterized by a Gaussian distribution with mean and variance parameters represented by $\mu$ and $\sigma$, respectively. The latent vector $z$ is then sampled from the dance latent space using a reparameterization technique~\cite{kingma2013auto} as input to the dance decoder. The dance decoder employs a cross-attention mechanism to fuse dance feature information and reconstruct the dance pose $\hat{x} \in \mathbb{R}^{L \times S}$. The cross-attention module utilizes zero tokens of dimension $L$ as the query matrix and the dance hidden vector as memory to ultimately reconstruct the dance sequence.

\subsection{Dance Diffusion}
Owing to the diversity and complexity of dances, the deployment of traditional generative networks is fraught. The mode collapse issue associated with GAN becomes pronounced, while the generative capability of flow models is constrained by bijections, rendering them ill-equipped to effectively handle high-diversity action generation tasks. The diffusion model, by virtue of its stochastic properties, is better suited for modeling dance movements characterized by high diversity distribution. The original dance sequence is replete with time-redundant information, and when used to construct a diffusion model directly, it impedes the generation effect and speed of the model. Consequently, we implement a denoising process on the dance latent space.

The diffusion model can be parametrically expressed as a Markov chain, with each step of the diffusion model equivalent to a transition process in the Markov chain. Our objective is to synthesize a coherent dance sequence $x^{1:L}$ that is stylistically consistent with the musical signal $c$. In contrast to the UNet network architecture~\cite{rombach2022high} previously employed in the image domain, we utilize a Transformer model to construct a dance diffusion generation framework that is better suited for continuous data such as audio sequences and dance motion sequences. 
The forward process of diffusion models is equivalent to gradually adding Gaussian noise to the original data distribution:

\begin{equation}
q\left(z_t^{1: L} \mid z_{t-1}^{1: L}\right)=\mathcal{N}\left(\sqrt{1-\beta_t} z_{t-1}^{1: L},\beta_t I\right)
\end{equation}

Where $z_0^{1: L}$ is drawn from the latent distribution, $\mathcal{N}(\cdot)$ denotes Gaussian distribution. The parameter $\beta_t$ are constant hyper-parameters for sampling, which is utilized to control the degree of noise addition. When $\alpha_t$ is closer to 1, $z_{T}^{1: L}$ can be approximated as a standard Gaussian distribution $\mathcal{N}(0, I)$. 



The reverse process of the diffusion model is equivalent to the gradual removal of noise from the Gaussian distribution. Each step is parameterized as follows:

\begin{equation}
p\left(z_{t-1}^{1: L} \mid z_{t}^{1: L}\right)=\mathcal{N}\left(z_{t-1}^{1: L}; \mu_\theta\left(z_{t}^{1: L}, t, c\right), \sigma_\theta\left(z_t^{1:L},t\right)\right) 
\end{equation}


Where $c$ represents the concatenated music features. $\sigma_\theta\left(x_t, t\right)$ is set to a time-related constant, which is equal to $\frac{1-\bar{\alpha}_{t-1}}{1-\bar{\alpha}_t} \beta_t$, where $\alpha_t=1-\beta_t$ and $\bar{\alpha}_t=\prod_{i=1}^t \alpha_i$.



This training process is parameterized by L2 loss, which is simplified by directly predicting the accumulated noise:

\begin{equation}
L(\theta)=\mathbb{E}_{z_0, \epsilon, t}\left[\left\|\epsilon-\epsilon_\theta\left(\sqrt{\bar{\alpha}_t} z_0+\sqrt{1-\bar{\alpha}_t} \epsilon, t, c\right)\right\|^2\right]
\end{equation}

where $t$ is sampled uniformly between 1 and $N$, $\epsilon \sim \mathcal{N}\left(0,1\right)$, and $\epsilon_\theta$ is the learned diffusion model.

\subsection{Genre Prediction Network}
Existing genre prediction methods are frequently hampered by a lack of generalization, attributable to the scarcity of large-scale music-dance datasets. To address this issue, we have incorporated the large-scale ChoreoSpectrum3D dataset into the Audio Spectrogram Transformer (AST) network~\cite{gong2021ast} for finetuning. Our architecture, inspired by the efficiency of LoRA-style fine-tuning, adapts pre-trained AST representations through learnable CNN-based skip connections. This design allows the model to synergize foundational audio features with domain-specific nuances from our ChoreoSpectrum3D dataset, ultimately extracting a 438-D stylistic vector for dance guidance.

Additionally, the AST network has been established on the foundation of ImageNet~\cite{russakovsky2015imagenet} and fine-tuned using training on the AudioSet dataset~\cite{hershey2017cnn}. To begin, we process the music waveform by transforming it into a 2D spectrogram. This spectrogram is partitioned into non-overlapping patch blocks, which are linearly projected into one-dimensional embeddings and augmented with learnable positional encodings. By fine-tuning this architecture on our extensive dataset, we ensure the contextual consistency of the generated dance motions.

\begin{figure}[htb]
    \centering
    \includegraphics[width=0.95\linewidth]{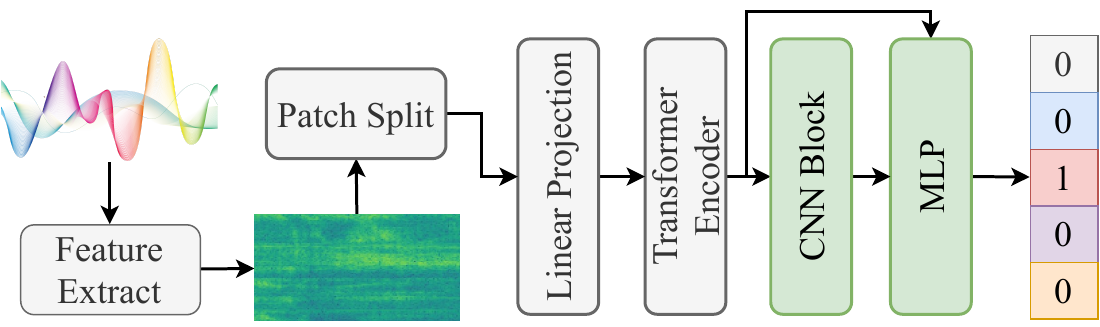}
    \caption{Genre Prediction Network}
    \label{fig:Genre-Prediction}
\end{figure}

The resulting patch sequence is then fed into a Transformer encoder, which generates high-level features for subsequent classification. This part of the network employs the pre-trained network and then performs fine-tuning when migrating to our dataset. Specifically, we introduce a simple CNN block to incorporate the output from the pre-training model. Utilizing skip connections, we combine this output with the output of the pre-training model. Finally, this combined output is passed through a final MLP classifier to predict the music genre category. The final music genre category is encoded using one-hot embedding, and the genre prediction network is optimized through supervised training using cross-entropy loss.
\begin{figure*}[htb]
    \centering
    \includegraphics[width=0.95\linewidth]{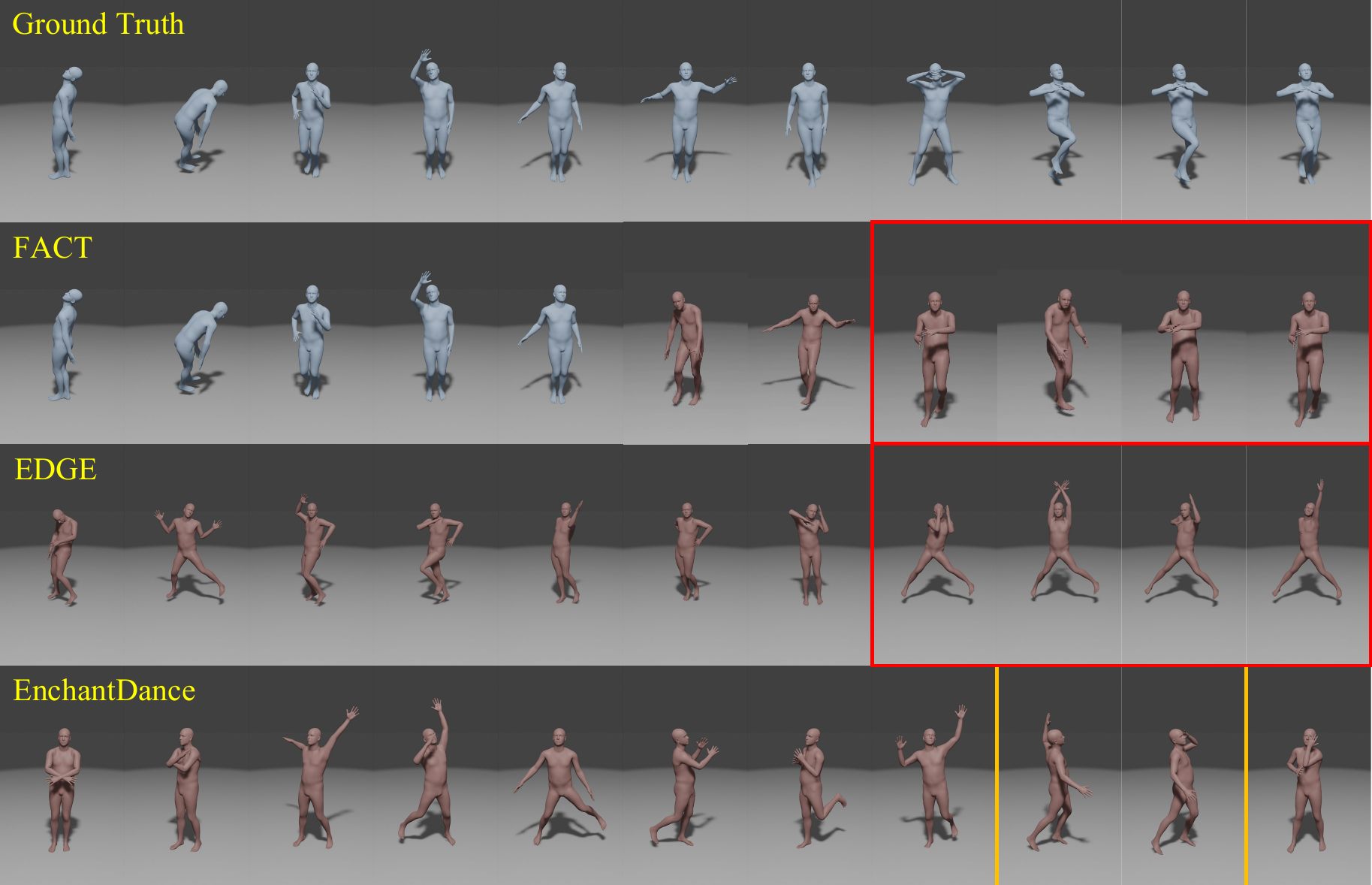}
    \caption{Visualization results. We conduct visual comparison (excluding Bailando) of the SMPL-based model, wherein FACT model employs an autoregressive method and necessitates seed motion input, with the blue portion representing the seed motion.}
    \label{fig:vis}
\end{figure*}

\section{Experiments}

\subsection{Dance Representation}
We represent dance motion sequences using two distinct forms: SMPL-based and HumanML3D-format. Both them have been demonstrated to be effective when utilized in our network. We mainly use SMPL parameters to conduct experiments, and the HumanML3D-format experiments can be found in Supplementary Material. For SMPL format, we represent dances as sequences of poses in the 24-joint SMPL pose parameters using the 3*3 Euler rotation matric, along with and translation matrix $X\in\mathbb{R}^{24\times 9+3=219}$.

\subsection{Evaluation Metrics}
Consistent with prior studies~\cite{siyao2022bailando,li2021ai}, we assess the generated dance motions based on three key aspects: quality, diversity, and consistency.

\subsubsection{Quality}
The quality of generated motion is evaluated using the Frechet Inception Distance (FID) \cite{heusel2017gans} to calculate the distribution distance between generated and ground-truth motions, as done in prior works \cite{siyao2022bailando}. Two well-designed motion feature extractors \cite{muller2005efficient, onuma2008fmdistance} were used to measure FID: (1) a kinetic feature extractor that maps a motion sequence $X$ to $z_k \in \mathbb{R}^{72}$, representing kinetic aspects of the motion such as velocity and acceleration. (2) a geometric feature extractor that produces a boolean vector $\mathbf{z}_g \in \mathbb{R}^{33}$ representing geometric relations between specific body points in the motion sequence $X \in \mathbb{R}^{T \times N \times 3}$. We denote the FID based on kinetic and geometric features as $\operatorname{FID}_k$ and $\operatorname{FID}_g^{\dagger}$, respectively.

\subsubsection{Diversity}
It evaluates the variations among generated dances corresponding to music inputs, reflecting the model’s generalization ability and its dependency on the input music. The average feature distance is used as the measurement, with features extracted using the same classifier employed in measuring FID, as done in prior works \cite{siyao2022bailando}. Motion diversity in the kinetic and geometric feature spaces is denoted as $\operatorname{Div}_k$ and $\operatorname{Div}_g^{\dagger}$, respectively.

\subsubsection{Consistency}
To evaluate the alignment between music and generated motions, the average temporal distance between each music beat and its closest dance beat is calculated as the Beat Align Score:

\begin{equation}
    \frac{1}{\left|B^m\right|} \sum_{t^m \in B^m} \exp \left\{-\frac{\min _{t^d \in B^d}\left\|t^d-t^m\right\|^2}{2 \sigma^2}\right\}
\end{equation}

where $B^d$ and $B^m$ record the time of beats in dance and music, respectively, while $\sigma$ is the normalized parameter.

\subsubsection{Physical Foot Contact score}
EDGE~\cite{tseng2023edge} introduces a metric for assessing physical plausibility, Physical Foot Contact score (PFC). This metric does not rely on explicit physical modeling but rather evaluates the rationality of physical motion by measuring the character’s acceleration and foot-ground contact. The underlying principle is that a character’s acceleration must result from static contact between the foot and the ground.

\begin{table}[h]
    \centering
    \scriptsize
    \caption{Evaluation on AIST++ Dataset. $\rightarrow$ indicates that the closer the value is to the ground truth, the better. \textcolor{red}{Red} and \textcolor{blue}{blue} indicate the first and the second best result seperately.}
    \label{tab: aist++}
    \begin{tabular}{cccccc}
    \toprule
    & \multicolumn{2}{c}{Motion Quality} & \multicolumn{2}{c}{Motion Diversity} & \\
     \cmidrule(lr){2-5}
    Method & $\operatorname{FID}_k \downarrow$ & $\operatorname{FID}_g^{\dagger} \downarrow$ &$\operatorname{Div}_k \rightarrow$ & $\operatorname{Div}_g^{\dagger} \rightarrow$ & BAS $\uparrow$   \\
    \midrule
    Ground Truth & 26.39 &12.21 & 6.13 &7.43 &0.2109  \\ \midrule
    FACT &62.19 &48.94 &14.73 &10.92 &0.1890  \\
    Bailando &\textcolor{blue}{46.35} &25.17 &11.98 &9.54 &0.1984 \\
    EDGE &91.82 &\textcolor{blue}{23.12} &\textcolor{blue}{8.73} &\textcolor{red}{7.18} &\textcolor{red}{0.2043}\\
    \midrule
    EnchantDance  &\textcolor{red}{28.09} &\textcolor{red}{7.56} &\textcolor{red}{5.39} &\textcolor{blue}{6.63} &\textcolor{blue}{0.1997} \\
    \bottomrule
    \end{tabular}
\end{table}

\begin{table}[h]
    \centering
    \scriptsize
    \caption{Evaluation on ChoreoSpectrum3D Dataset}
    \label{tab: ChoreoSpectrum3D}
    \begin{tabular}{cccccc}
    \toprule
    & \multicolumn{2}{c}{Motion Quality} & \multicolumn{2}{c}{Motion Diversity} & \\
     \cmidrule(lr){2-5}
    Method & $\operatorname{FID}_k \downarrow$ & $\operatorname{FID}_g^{\dagger} \downarrow$ &$\operatorname{Div}_k \rightarrow$ & $\operatorname{Div}_g^{\dagger} \rightarrow$ & BAS $\uparrow$   \\
    \midrule
    Ground Truth &3.31 &0.09 & 8.45 &6.74 &0.2362  \\ \midrule
    FACT  &75.36 &69.35 &10.73 &21.21 &0.2110   \\
    Bailando  &\textcolor{blue}{50.23} &54.85 &7.49 &19.61 &0.2253  \\
    EDGE  &98.02 &\textcolor{blue}{53.20} &\textcolor{red}{8.72} &\textcolor{blue}{8.81} &\textcolor{red}{0.2483}  \\
    \midrule
    EnchantDance  &\textcolor{red}{4.80} &\textcolor{red}{11.38} &\textcolor{blue}{8.08} &\textcolor{red}{7.42} &\textcolor{blue}{0.2421}\\
    \bottomrule
    \end{tabular}
\end{table}

\subsection{Implementation Details}

\textbf{Motion Compression and Dance VAE.} To handle this scale efficiently and mitigate the motion redundancy inherent in direct diffusion on raw data, we emphasize the necessity of motion compression. The Dance VAE employs the Transformer architecture, comprising 9 layers and 4 heads in both the encoder and decoder, supplemented by residual connections. This design follows the observation that compressing dance motions into dense latent features is essential for maintaining high generative quality when dealing with large-scale datasets.

\textbf{Music Representation.} Music features are extracted using Jukebox~\cite{dhariwal2020jukebox} and downsampled to match the motion frame rates of 60 FPS for AIST++ and 20 FPS for ChoreoSpectrum3D. The music representation, with an initial feature dimension of 4800, is encoded to 512 dimensions through a two-layer Transformer encoder. 

\textbf{Training and Inference}
The model is trained using the AdamW optimizer with a fixed learning rate of $10^{-4}$. For the diffusion process, we set the number of steps to 1,000 during training and 50 for inference. The variances $\beta_t$ are linearly scaled from $8.5 \times 10^{-4}$ to 0.012. Following standard protocols, we re-trained existing baselines like FACT and Bailando from their released checkpoints or from scratch until convergence to ensure a fair comparison.

\subsection{Comparision and Results}
We conduct comparative experiments of several existing state-of-the-art methods, including FACT, Bailando, and EDGE, on the AIST++ and ChoreoSpectrum3D datasets. Given the disparate data division methods employed in the AIST++ dataset, we uniformly adopt a sliding window of 240 frames with a step size of 40 frames to obtain the final samples, yielding a total of 20,785 data samples with an action FPS of 60. For each method, we employ the identical data partition strategy. The quantification results on the AIST++ dataset are presented in Table~\ref{tab: aist++}. For the FACT model, we finetune the release checkpoint until convergence is achieved. For the EDGE model, which downsamples the original data to 30 FPS, we modify it to 60 FPS to maintain a fair comparison. For the Bailando model, we employ the default parameters from the open-source code. 

According to the Table~\ref{tab: aist++}, EnchantDance outperforms the best baseline model by 18.26 (39\%) and 15.56 (67\%) on $\operatorname{FID}_k$ and $\operatorname{FID}_g^{\dagger}$, respectively. These two values represent the physical and manual template features of the dance, respectively, directly reflecting the improvement in dance quality. 
EnchantDance also achieves the best results on $\operatorname{Div}_k$, indicating that our method is capable of generating physically diverse dances. Although our method is only 0.55 (7\%) worse than the best EDGE on $\operatorname{Div}_g^{\dagger}$, a comprehensive assessment of the improvement effect of $\operatorname{FID}_g^{\dagger}$ reveals that our method can still keep pace with diversity while maintaining the best generation quality. Our method is slightly inferior to EDGE with respect to BAS. 

We also train and evaluate the SOTA method on the ChoreoSpectrum3D Dataset, with the quantitative results presented in Table~\ref{tab: ChoreoSpectrum3D}. These results largely corroborate the conclusions drawn above. According to Tables~\ref{tab: aist++} and~\ref{tab: ChoreoSpectrum3D}, while EDGE does not perform as well as other methods in $\operatorname{FID}_k$, it frequently holds a leading position in BAS. Figure~\ref{fig:vis} displays the visualization results, which demonstrate that the dance movements generated by EDGE are often relatively large, thus facilitating the detection of dance beats and contributing to an improvement in BAS. The red box in Figure~\ref{fig:vis} indicates the number of frames with poor performance. It can be observed that EDGE often has a foot without touching ground, leading to its suboptimal performance on $\operatorname{FID}_k$, which primarily measures the motion's physical plausibility. 

EDGE also introduces PFC to measure the physics plausibility. As shown in Table~\ref{tab: PFC}, our results on the PFC are superior to those of the EDGE method, further confirming that EnchantDance can generate physically plausible dances. For the FACT model, it can be seen that due to the limitations of the autoregressive model, its limbs are in a stiff state in the second half of the dance. The FACT original text only uses 120 frames of seed motion as input and predicts 20 frames of data, which is insufficient for generating realistic long-term dances. For the yellow part, it can be observed that the dance movements corresponding to the ground truth have not changed significantly and can thus be considered to be within the same dance beat. The corresponding frames of our method and the EDGE method exhibit similar action changes, providing evidence that both EnchantDance and EDGE methods are leading in BAS.

\begin{table}[h]
    \centering
    \scriptsize
    \caption{Evaluation on Physical Foot Contact score.}
    \label{tab: PFC}
    \begin{tabular}{cccc}
    \toprule
    Dataset & EDGE & EnchantDance &Ground Truth   \\
    \midrule
    AIST++ & 1.1865 &\textcolor{red}{0.8006} & 0.7305   \\ 
    ChoreoSpectrum3D & 14.8483 &\textcolor{red}{4.2182} &7.5227 \\
    \bottomrule
    \end{tabular}
\end{table}

\subsection{In-the-Wild Music}

The ability to generate generalizable dance motions is essential. To assess generalization, we implement a bidirectional cross-dataset evaluation: (i) generating dances for ChoreoSpectrum3D music using an AIST++-trained model, and (ii) vice versa. This scheme effectively tests the model's robustness on out-of-distribution audio-motion pairs.

For our experiments, we select 1000 music samples from each dataset. Specifically, for the AIST++ dataset, we select 100 samples for each of its ten dance genres. For the ChoreoSpectrum3D dataset, we select 250 samples for each of its four dance genres. The results are shown in Table~\ref{tab: generalize_aist++} and Table~\ref{tab: generalize_ChoreoSpectrum3D}. Our method outperforms other methods on out-of-set data, while the collapse of the Bailando model is more obvious. To evaluate the generalization capability of EnchantDance in real-world scenarios, we conducted "in-the-wild" testing using four popular tracks across diverse genres: Woman (Doja Cat), Despacito (Luis Fonsi), LOCO (ITZY), and My Type (Saweetie), following EDGE~\cite{tseng2023edge}.

\begin{table}[h]
    \centering
    \scriptsize
    \caption{Generalize evaluation on AIST++ Dataset (employ model trained on ChoreoSpectrum3D Dataset)}
    \label{tab: generalize_aist++}
    \begin{tabular}{cccccc}
    \toprule
    & \multicolumn{2}{c}{Motion Quality} & \multicolumn{2}{c}{Motion Diversity} & \\
     \cmidrule(lr){2-5}
    Method & $\operatorname{FID}_k \downarrow$ & $\operatorname{FID}_g^{\dagger} \downarrow$ &$\operatorname{Div}_k \rightarrow$ & $\operatorname{Div}_g^{\dagger} \rightarrow$ & BAS $\uparrow$   \\
    \midrule
    Ground Truth &4.93 &0.42 &9.13 &7.41 &0.2130   \\ \midrule
    Bailando  &394961.25 &105.53 &308.13 &\textcolor{blue}{9.24} &0.2321 \\
    EDGE  &\textcolor{blue}{2511.38} &\textcolor{blue}{35.58} &\textcolor{blue}{29.16} &4.41 &\textcolor{red}{0.2452} \\
    \midrule
    EnchantDance  &\textcolor{red}{114.65} &\textcolor{red}{20.91} &\textcolor{red}{12.89} &\textcolor{red}{7.51} &\textcolor{blue}{0.2387} \\
    \bottomrule
    \end{tabular}
\end{table}

\begin{table}[h]
    \centering
    \scriptsize
    \caption{Generalize evaluation on ChoreoSpectrum3D Dataset (employ model trained on AIST++ Dataset)}
    \label{tab: generalize_ChoreoSpectrum3D}
    \begin{tabular}{cccccc}
    \toprule
    & \multicolumn{2}{c}{Motion Quality} & \multicolumn{2}{c}{Motion Diversity} & \\
     \cmidrule(lr){2-5}
    Method & $\operatorname{FID}_k \downarrow$ & $\operatorname{FID}_g^{\dagger} \downarrow$ &$\operatorname{Div}_k \rightarrow$ & $\operatorname{Div}_g^{\dagger} \rightarrow$ & BAS $\uparrow$   \\
    \midrule
    Ground Truth &1.27 &2.12 &8.79 &6.98 &0.2357  \\ \midrule
    Bailando  &\textcolor{blue}{366.63} &574.32 &\textcolor{blue}{1.98} &25.52 &0.1957 \\
    EDGE  &412.25 &\textcolor{blue}{338.54} &0.92 &\textcolor{blue}{13.07} &\textcolor{blue}{0.2351} \\
    \midrule
    EnchantDance  &\textcolor{red}{16.61} &\textcolor{red}{21.40} &\textcolor{red}{7.61} &\textcolor{red}{7.85} &\textcolor{red}{0.2415} \\
    \bottomrule
    \end{tabular}
\end{table}

An interesting observation arises from our cross-dataset evaluation in Table~\ref{tab: generalize_aist++} and Table~\ref{tab: generalize_ChoreoSpectrum3D}. When trained on ChoreoSpectrum3D, the model generates a diverse range of motions that far exceed the distribution of the smaller AIST++ dataset, leading to artificially inflated FID scores when AIST++ is used as the reference. Conversely, models trained on AIST++ remain within the distribution of ChoreoSpectrum3D, avoiding this issue. This phenomenon highlights a limitation in the current music-to-dance FID metric: it tends to penalize models trained on superior, larger-scale datasets for capturing a broader motion distribution that the smaller reference dataset cannot represent.

As the dance movements generated by wild music do not provide FID and diversity evaluation values, we conduct a quantitative evaluation on the BAS and PFC metrics. The evaluation results are presented in Table~\ref{tab:wild}. As can be observed, the generalization ability of the model trained on the ChoreoSpectrum3D dataset significantly surpasses that of the model trained on the AIST++ dataset. Furthermore, we conduct a comparison with the EDGE model, which further verifies the superior generalization ability of our EnchantDance model. 

\begin{table}[htp]
    \centering
    \caption{Evaluation on wild Youtube music}
    \label{tab:wild}
    \begin{tabular}{c|c|c}
    \toprule
     Method  &BAS $\uparrow$ & PFC $\downarrow$ \\ \midrule
     EDGE w/ AIST++ &0.2169 &2.2563\\
     EDGE w/ ChoreoSpectrum3D &\textcolor{blue}{0.2386} &1.5437\\
     EnchatDance w/ AIST++ &0.2102 &\textcolor{blue}{1.5108}  \\
     EnchatDance w/ ChoreoSpectrum3D &\textcolor{red}{0.2467}&\textcolor{red}{0.6542 } \\
     \bottomrule
    \end{tabular}
\end{table}

\subsection{Ablation Study}
We conducted two ablation studies: 1) without utilizing the music genre feature, and 2) directly using the feature representation from the last layer of the AST model as the music genre feature, concatenated with Jukebox features. Keeping the weights of the VQ-VAE unchanged, each model was trained on a single Tesla A100 GPU for 200 epochs, taking approximately 48 hours, and the evaluation results are shown in Table~\ref{tab: ChoreoSpectrum3D}. The fine-tuned music genre features achieved top-1 performance across all metrics. Additionally, the visualization results are depicted in Fig.~\ref{fig: genre}, where yellow boxes indicate ballet dances and red boxes indicate pop dances. The first row represents w/ genre, and the second row represents w/o genre; they each correspond to dance sequences generated from the same piece of music. We selected frames at identical timestamps for comparative analysis. It is evident that the iconic spinning movements in ballet are pronounced in the first row within the yellow box, and similarly, the characteristic side-step initiation movements in pop dances are prominent in the first row within the red box.

\begin{table}
    \centering
    \scriptsize
    \caption{Ablation experiment evaluation}
    \label{tab: ChoreoSpectrum3D}
    \begin{tabular}{cccccc}
    \toprule
    Method & $\operatorname{FID}_k \downarrow$ & $\operatorname{FID}_g^{\dagger} \downarrow$ &$\operatorname{Div}_k \rightarrow$ & $\operatorname{Div}_g^{\dagger} \rightarrow$ & BAS $\uparrow$   \\
    \midrule
    w/o genre &6.54 &12.32 &7.96 &7.63 &0.2401  \\ 
    w/ AST &6.27 &12.07 &7.89 &7.71 &0.2411 \\ \midrule
    EnchantDance  &\textbf{4.80} &\textbf{11.38} &\textbf{8.08} &\textbf{7.42} &\textbf{0.2421}\\
    \bottomrule
    \end{tabular}
\end{table}

\begin{figure}
    \centering
    \includegraphics[width=1.0\linewidth]{PIC/ballet.pdf}
    \caption{Ablation study on genre-aware features (Ballet example). A visual comparison of generated dance movements with and without genre conditions. }
    \label{fig: genre}
    \vspace{-0.6cm}
\end{figure}

\section{Conclusion}

In this paper, we propose EnchantDance, a latent-based generative framework integrated with a genre prediction network to synthesize high-fidelity dance sequences that align seamlessly with musical rhythms. Extensive evaluations demonstrate that EnchantDance outperforms state-of-the-art methods in generating realistic and diverse motions, particularly exhibiting superior generalization on out-of-distribution musical data. Furthermore, we contribute a large-scale, high-quality music-dance dataset totaling 70.32 hours. As the most extensive synchronized music-dance collection to date, this dataset serves as a critical foundation for advancing model robustness and motion quality. Although the current iteration of EnchantDance does not explicitly support interactive motion editing, its latent architecture possesses significant generative potential. Future research will focus on extending the framework to support dance editing and a broader spectrum of dance genres.

\bibliographystyle{IEEEtran}
\bibliography{reference}

\end{document}